# Electrothermal manipulation of current-induced phase transitions in ferrimagnetic $Mn_3Si_2Te_6$

Jiaqi Fang[1,2,#], Jiawei Hu[1,2,#], Xintian Chen[1,2,#], Yaotian Liu[1,2], Zheng Yin[1,2], Zhe Ying[1], Yunhao Wang[1,2], Ziqiang Wang[3], Zhilin Li[1,2,*], Shiyu Zhu[1,2,*], Yang Xu[1,2,*], Sokrates T. Pantelides[4,2], and Hong-Jun Gao[1,2,5,*]

[1]Beijing National Laboratory for Condensed Matter Physics, Institute of Physics, Chinese Academy of Sciences, Beijing 100190, China

[2]School of Physical Sciences, University of Chinese Academy of Sciences, Beijing 100049, China

[3]Department of Physics, Boston College, Chestnut Hill, MA, 02467, USA

[4]Department of Physics and Astronomy & Department of Electrical and Computer Engineering, Vanderbilt University, Nashville, TN 37235, USA

[5]Hefei National Laboratory, 230088 Hefei, Anhui, China.

*Correspondence to: lizhilin@iphy.ac.cn, syzhu@iphy.ac.cn, yang.xu@iphy.ac.cn, hjgao@iphy.ac.cn

#These authors contributed equally: Jiaqi Fang, Jiawei Hu, Xintian Chen

**Abstract**

**Phase transitions driven by external stimuli are central to condensed matter physics, providing critical insights into symmetry breaking and emergent phenomena. Recently, ferrimagnetic (FiM) $Mn_3Si_2Te_6$ has attracted considerable attention for its magnetic-field-induced insulator-metal transitions (IMTs) and unconventional current-driven phase transitions, yet the role of applied currents in the magnetic phase remains poorly understood. Here, by combining local magnetization probes and time-resolved transport measurements, we uncover an electrothermal origin for the current-induced first-order-like phase transitions, characterized by abrupt voltage jumps and distinct magnetic domain evolution. Current-voltage (*I-V*) characteristics measured under triangular waveforms exhibit strong non- reciprocal and hysteretic behaviors, which are significantly suppressed at frequencies ~1000 Hz. Time-resolved studies using rectangular pulsed currents demonstrate that the resistance dynamics closely mirror the equilibrium resistance-temperature profile, directly implicating Joule heating as the driving mechanism. Furthermore, we reveal that the intrinsic *I-V* response adheres to Ohm's law, displaying linearity across various magnetic fields and temperatures. Our work advocates for a cautious approach in distinguishing between genuine current-induced nonequilibrium quantum states and thermal effects.**

**Introduction**

Electric currents or fields can introduce emergent nonequilibrium quantum states, such as the electromagnetic radiation-induced superconductivity[1–3] and electrically driven IMTs (e.g., VO$_2$, Ca$_2$RuO$_4$, Sr$_2$IrO$_4$)[4,5]. These transitions are often accompanied by sharp resistance change, negative differential resistance (NDR), and bistable switching, making them suitable candidates for logic and memory electronics[6–8]. However, in many cases, unambiguous evidence for the genuine non-equilibrium steady states after the transitions remains elusive, largely due to limitations in detection methods and the intricate interplay of various mechanisms, including unavoidable Joule heating that drives the system into higher-temperature phases[4,9–11]. Additionally, some materials simultaneously exhibit electronic, magnetic, and structural phase transitions, further complicating the analysis[12,13]. Despite numerous efforts, experimentally distinguishing and precisely controlling electro-thermal effects from nonthermal origins are still challenging[13–17].

Mn$_3$Si$_2$Te$_6$ has recently emerged as a novel layered ferrimagnet (FiM) with alternating stacking of honeycomb and triangular lattices of Mn atoms, whose spins are predominantly lying within the *ab* plane (Fig. 1a)[18–21]. An out-of-plane magnetic field can drive IMTs, manifesting as colossal magnetoresistance (CMR) with strong anisotropic field dependences[18–24]. The spin orientation and its coupling to the orbital degree of freedom have been proposed to control the splitting of topological nodal lines and give rise to an efficient gap closing near the Fermi level[21,23,25,26]. In another work, Y. Zhang *et al.* reported the observation of peculiar current-induced transport behaviors, i.e., first-order-like phase transitions and time-dependent bistable switching, as hallmarks of the existence of chiral orbital currents (COC) that can couple to the ferrimagnetic state and flow spontaneously along edges of MnTe$_6$ octahedra in Mn$_3$Si$_2$Te$_6$[27].

Here we report a study combining current-modulated local magnetization microscopy and dynamical current-voltage measurements on bulk samples of Mn$_3$Si$_2$Te$_6$ to scrutinize the effect of current flow. Using MFM integrated with electrical transport capabilities, we observe a current-driven magnetic phase transition via evolution of magnetic domains. The switching between FiM and paramagnetic (PM) phases exhibit distinct hysteretic behavior, concomitant with a sudden jump in resistance. Transport measurements with frequency-tunable triangular waves reveal similar hysteretic behavior at low frequencies. However, the charge transport demonstrates significant nonreciprocity, with polarity determined by the direction of current sweeping and amplitude strongly dependent on the frequency. Additionally, pulsed current transport measurements with temporal resolution further reveal that the detailed resistance evolution closely mirrors the resistance-temperature (*R-T*) profile. We highlight that under tens of mA currents the real sample temperature ($T_s$) can be raised by >100 K in less than 1 ms. Such self-heating drives the sample into a higher-temperature nonmagnetic and semiconducting phase, consistently explaining our observations. We further demonstrate the intrinsic current-voltage relation of $Mn_3Si_2Te_6$ to have linear characteristics.

## Results

### Visualization of the magnetic ground states

To investigate the microscopic magnetic characteristics, we first conducted temperature-dependent MFM measurements on a $Mn_3Si_2Te_6$ sample. The cryostat is equipped with a three-axis superconducting vector magnet that can help distinguish spin orientations of the magnetic domains. The shift of resonance frequency $\Delta f$ of the tip (magnetized vertically) is proportional to the second derivative of the out-of-plane stray field (emanating from the sample) with respect to the tip height. At $T = 25$ K (far below Curie temperature $T_c$) and zero magnetic field ($H = 0$), the in-plane FiM domains manifest as bright regions in the MFM mapping, separated by a transverse dark domain wall (Fig. 1c). These FiM domains shrink and expand easily under small in-plane magnetic fields in the $Y$ direction, suggesting their magnetization in parallel/antiparallel with the domain wall (Supplemental Fig. 1). The profile of $\Delta f$ shows prominent dip and nearly symmetric values in the profile crossing two domains, indicating a Bloch-type domain wall as schematically illustrated in Fig. 1c[28–30].

The evolution of magnetic domains with increasing the temperature is shown in Fig.1d. As the sample warms up through $T_c \sim 83$ K (slightly higher than previously reported[19]), the width of the domain wall expands and the sample evolves from FiM to the PM phase at $T > T_c$. The dynamic process of domain wall expansion (contraction) during warming (cooling) is observed in detailed temperature dependent MFM measurements (Supplemental Fig. 2). A quantitative evolution of the magnetic state is revealed by the extracted PM ratio obtained from analyzing the $\Delta f$ distribution of the FiM domain wall (Fig. 1e). A dramatic change of FiM region ratio around $T_c \sim 82.8$ K is observed, indicating a temperature-driven magnetic phase transition. This observation is consistent with the global magnetization measurements performed along the $ab$ plane (inset of Fig. 1e).

We next study the current-driven phase transition[27] by supplying direct-current flow through the sample along with the MFM measurements. The MFM mapping reveals a FiM state at low currents, indicated by domain and domain wall signals, and a nonmagnetic state at high currents, with no detectable MFM contrast (Fig. 1f). To perform continuous MFM measurements, the tip is maintained at a constant height above the sample and scanned across the domain wall, while the direct current is swept upwards or downwards. During the current-ramping-up process, the domain wall signal first persists and barely changes, suggesting a robust FiM state. However, when the current surpasses a threshold or critical value of $I_{c\uparrow} \sim 17.4$ mA, we observe a significant change in MFM signals and a sudden disappearance of the domain wall (Fig. 1g, left). The magnetic features observed beyond the critical currents are similar to those of the PM phase above $T_c$ in Fig. 1d. This behavior indicates that a nonmagnetic phase emerges at larger currents. Similarly, during the current-reduction process, the FiM state is recovered below a smaller critical value of $I_{c\downarrow} \sim 15.6$ mA, showing hysteretic behavior (Fig. 1g, right). The quantitative evolution of the current-dependent magnetic state is further revealed by the root-mean-square of the MFM line scan ($\Delta f_{rms}$) (Fig. 1h).

The four-terminal $I$-$V$ curves, measured simultaneously with the MFM mappings, exhibit strong nonlinearity and identical hysteretic transitions at the critical currents $I_{c\uparrow}$ and $I_{c\downarrow}$ (Fig. 1i). The slope of the $I$-$V$ relationship is small under low currents, followed by a relatively broad region where the voltage changes gradually before reaching the critical current. The abrupt change of the voltage signals at the transition and the resulting NDR ($dV/dI<0$) at $I > I_c$ are reminiscent of the previously reported first-order and bistable switching observed for Mn$_3$Si$_2$Te$_6$[27], as well as the electric-field-driven IMTs observed in many transition-metal compounds[4]. It is hence crucial to investigate the origins of the current-induced switching and the NDR associated with the nonmagnetic phase $I > I_c$. Given that the resistance is well above a kiloohm (kΩ) at low temperatures, a current above a milliamp (mA) could induce substantial Joule heating that cannot be safely neglected[4,31]. Therefore, the electro-thermal effects need to be carefully scrutinized. In the following, we utilize fast time-resolved transport measurements in order to achieve precise control and detection of the first-order-like switching[10,15,32–34].

**Dynamical current-voltage responses**

In typical transport measurements, DC or small-frequency AC measurements with time resolution ~100 ms for the data acquisition were employed to obtain the electrical responses (Supplemental Fig. 3b), where any transient information associated with the fast current-driven transitions cannot be well resolved. To improve the temporal resolution to microseconds (μs), we performed measurements using an experimental setup shown in Fig. 2a. The current $I$ is generated by a source meter (Keithley 6221) that can produce various waveforms (mainly triangular and rectangular waves used here), flowing through the Mn$_3$Si$_2$Te$_6$ sample and a standard resistor ($R_0 = 10$ Ω). The corresponding voltages ($V_i$, i=1, 2, 3, and 4) at different probes are simultaneously recorded by a 4-channel digital oscilloscope. The transient four-terminal voltage and current through the sample can then be obtained by $V = V_2 - V_3$ and $I = V_4/R_0$, respectively.

We first manipulate the current-driven phase transition by applying triangle waves with peak current ±35 mA and varying the repetition rates $f$ at a cryostat temperature $T$ = 60 K (cryogenic bath temperature read by the thermometer) for a different $Mn_3Si_2Te_6$ sample. A schematic oscillogram in Fig. 2a shows the detected $I$ and $V$ traces with slowly varying currents at $f$ = 0.1 Hz. In Fig. 2b, we present the data in the form of $I$-$V$ curves for the upstream (forward) and downstream (backward) directions (color-coded, denoted by the orange and green arrows in Fig. 2a, respectively) at a few representative frequencies. As can be seen, the $I$-$V$ curves display strong frequency-dependent shapes and hysteretic behaviors. At low frequencies (0.1 Hz, for example), the two curves coincide with the ones measured under static DC currents (Supplemental Fig. 3b), exhibiting linear $I$-$V$ behavior at small currents, followed by a sharp transition at a critical current $I_c$ of ~7 mA (slightly different for orange and green curves), and then showing NDR. As the frequency increases, the discrepancy between the forward- and backward-swept curves becomes more pronounced, reaching its maximum near 100 Hz. The two curves are related by a sign reversal of $V \to -V$ and $I \to -I$. While for each curve, the antisymmetry is lost, namely $V(I) \neq -V(-I)$ or $R(I) \equiv \frac{V(I)}{I} \neq R(-I)$, indicating strong nonreciprocity and non-Ohmic behavior. At higher frequencies (>~1000 Hz), the $I$-$V$ curves tend to become linear in the whole current range.

We define the degree of nonreciprocity by $\eta = \frac{R(I)-R(-I)}{R(I)+R(-I)}$ to quantify the magnitude of asymmetry of the resistance derived from each $I$-$V$ curve[35]. The bidirectional $\eta$ is plotted in Fig. 2c (calculated from data in Fig. 2b) and Fig. 2d (full frequency range). The closer $\eta$ is to 0, the smaller the discrepancy between the forward- and backward-swept $I$-$V$ curves and the higher the symmetry of each curve. At low frequencies, $\eta$ deviates appreciably from 0 only near $I_c$, stemming from the difference in $I_c$ when sweeping the current forward and backward. The $I_c$ decreases with increasing $f$, and disappears near a critical frequency $f_c$ ~ 70 Hz. At higher frequencies, $\eta$ reaches a maximum value of ~0.4 near 100 Hz at $I$ = ~15 mA and approaches 0 again at 1000 Hz. Phenomenologically, the deviation of $\eta$ from 0 implies nonreciprocal and diode-like transport behavior. However, in conventional conductors, a non-zero $\eta$ is typically rare and extremely difficult to be directly observed in DC transport measurements[36]. The abnormally large $\eta$ and the dependences on its frequency and sweeping direction observed here are puzzling at first sight. We have also extracted the differential resistance (d$V$/d$I$) near zero bias voltage at various frequencies (Fig. 2e) and observed a nonmonotonic frequency dependence. It has a broad peak near $f_c$ ~ 70 Hz and tends to saturate at low and high frequencies. The strong frequency dependences of both $\eta$ and d$V$/d$I$ ($V$ = 0) indicate slow processes happening in the sample. We notice that the power supplied by the AC current at 1000 Hz is approximately 0.13 W. Considering the sample heat capacity $C$ (calculated by the specific heat from the literature[19,24] and the estimated amount of substance of the sample), the Joule heating could raise the sample temperature by ~105 K if dissipation in a short time scale, ~1 ms, is neglected. Further examination with short current pulses would help reveal the role of Joule heating more clearly.

In Fig. 3a, we show the $R$-$T$ curve in thermal equilibrium measured at small excitation currents (100 μA used here), consistent with previous measurements[19,22,23]. The sample exhibits thermal-activated semiconducting behavior (d$R$/d$T$<0) at both higher temperatures (>~$T_c$) and lower temperatures (<~50 K), with a metallic behavior (d$R$/d$T$>0) at intermediate temperatures. We then investigate the sample under pulsed currents (rectangular waveform, see a schematic in Fig. 3b) with an onset duration of 500 μs and an amplitude of $I_a$, followed by a delay time of ~5 s (repetition rate $f$ = 0.2 Hz) to avoid any accumulative heating.

We now focus on the electrical responses within the 500 μs current pulse. The measurements are performed with varying $I_a$ at cryostat temperatures $T$ of 20, 50, 70, 100 K (color-coded and highlighted by the circles in the $R$-$T$ curve shown in Fig. 3a), with an example of the real-time voltage response at $T$ = 20 K shown in Fig. 3c. As demonstrated at $T$ = 20 K, the response strongly depends on the current pulse amplitude, varying from 6 mA to 40 mA. At small $I_a$, it features a sharp initial rise and a long decay afterward. With increasing $I_a$, the decay in $V$ slowly evolves into an upturn (showing a broad dip) and then a peak feature emerges at large currents. The larger $I_a$ results in faster onsets of the minima (marked by the star symbols) and maxima (marked by the pentagon symbols). Combining the measured real-time current $I$, we can obtain the time dependences of the resistance with $R$ = $V$/$I$ in Fig. 3d (also shown here for the results at $T$ = 50 K, 70 K, and 100 K). Interestingly, as illustrated by the orange (green) horizontal dashed lines in the top panel of Fig. 3d, the minima (maxima) collapse to the same resistance value, ~220 Ω (~600 Ω) after the scaling. The initial rise of the resistance curves within a few microseconds is attributed to the charging of the parasitic capacitance (estimated to be on the order of ~100 pf) in the cryogenic circuits.

While the evolution of the resistance with respect to time at 50 K and 70 K does not show the dip feature, the resistance maxima still exist, exhibiting a trend of shorter onset time with increases in $I_a$ and $T$. The values of the resistance maxima increase slightly compared to those at 20 K. As the temperature rises to 100 K, the time-dependent resistance only exhibits a long-decaying feature that goes to lower values at larger $I_a$. We also notice that the real-time resistance evolution measured at the four different cryostat temperatures (20, 50, 70, or 100 K) with large currents (e.g. the orange curves) in Fig. 3d, closely resembles the segments of the $R$-$T$ curve at higher temperatures (than that marked by the colored dot symbols) in Fig. 3a, indicating a strong correlation between time progression and temperature increase in the pulsed-current measurements.

In the following analysis, we will show that the cumulative heating effect produced by the currents can well explain the experimental observations. We employ a simplified electro-thermal model simulation for further examination. The sample temperature $T_s$ (assumed to be uniform for simplicity) is governed by the balance between Joule heating and thermal dissipation, described by: $C\frac{dT_s}{dt} = I^2 R - K(T_s - T_0)$, where $C$ is the sample heat capacity, $K$ is the interfacial thermal conductance (see more details in methods). We assume a constant cryostat and starting temperature $T = T_0$ (read by the thermometer) during the short-time measurements. The parameters $C$, $K$, and $R$ are temperature dependent. To approximate the complexity of heat exchange between the sample and its environment, $K$ is modeled using a second-order polynomial fit [37]. For the resistance $R$, we simply adopt Ohm's law $R = V/I$ without considering any dependence on the applied current (data taken from Fig. 3a). Using this framework, we simulate the time evolution of $R$ for varying pulse amplitudes $I_a$ (e.g., simulated results at $T = 20$ K in Fig. 3e) and the corresponding increase in $T_s$ (Fig. 3f). The simulated $R$-$t$ curves reproduce the key features observed experimentally (top panel of Fig. 3d). Upon current application ($t = 0$), Joule heating drives $T_s$ upward, and $R$ evolves according to the temperature-dependent $R$-$T$ curve. Higher currents induce faster heating and larger temperature increases. The simulated onset times for resistance peaks and dips (white curves in Fig. 3g) align with experimental data (symbols), as visualized in the resistance color map.

The simulation indicates that the sample temperature can rise rapidly by more than ~100 K within the short current pulse (far off from the cryostat temperature $T$ read by the thermometer). We note that the resistance peaks in Fig. 3d are slightly broader and have smaller values than the maximal resistance ~860 Ω in Fig. 3a. This should be due to the inhomogeneous heating in the bulk sample, as can be seen from the temperature distribution simulated using COMSOL in the inset of Fig. 3f. The dynamical $I$-$V$ characteristics under triangular waveforms can also be simulated using the Joule-heating model with the same parameters (Supplemental Fig. 4), revealing similar hysteretic features as that shown in Fig. 2b.

To rule out an impact by Joule heating on the $I$-$V$ curves, we further validate our analyses by extracting the intrinsic $I$-$V$ characteristics from the pulsed-current measurements (Figs. 4a and 4b). The sample voltage $V$ and current $I$ are recorded shortly after the onset of the current pulse (delay time $t = 7.5$ μs used here) when no substantial Joule heating is involved. We note that such methods have been extensively used for obtaining the intrinsic $I$-$V$ characteristics under large electric fields or currents[31,33,34]. Here, the obtained $I$-$V$ curves for $Mn_3Si_2Te_6$ exhibit linear relationships at all the measured magnetic fields and temperatures, *i.e.*, they obey Ohm's law. This result is in direct contrast to the slow DC measurements that show critical-current-like behaviors (see Supplemental Fig. 3b).

In the upper panel of Fig. 4c, we directly compare the *I-V* curves derived from pulsed measurements (connected purple symbols) with DC transport measurements (black curve) at 20 K as an example. The two curves overlap closely in the low current regime ($I<$ ~1 mA) but diverge significantly at higher currents. Notably, the sharp transition and NDR observed in the DC curve are absent in the intrinsic *I-V* curve. The resistance $R$ ($= V/I$) calculated from the DC *I-V* curve (lower panel of Fig. 4c), exhibits a nonmonotonic dependence on current, mirroring its temperature-dependent evolution over the range ~20 K$<T<$~140 K (Fig. 4d). Regions of correspondence between the current- and temperature-dependent $R$ are color-coded (blue, pink, orange). A key distinction arises in the pink regions, where the resistance increases at the critical current $I_c$~13 mA is markedly sharper than the analogous temperature-driven rise in the range 50 K$<T<$90 K. This discrepancy is attributed to Joule heating and an associated positive-feedback mechanism. During DC measurements, initial current increases gradually heat the sample to 50 K, resulting in a negative *R-I* dependence (blue region) due to the intrinsic decrease of resistance with temperature ($dR/dT<0$). Beyond this point, further current increases raise the sample temperature ($T_s$), driving $R$ upward. This enhances the Joule power ($P = IV = I^2R$), which further promote a rise in $T_s$ through the positive feedback loop (Fig. 4d, inset). Consequently, the *R-I* curve exhibits a steeper rise at $I_c$ compared to the *R-T* behavior, reflecting the self-reinforcing interplay between heating and resistance.

The temperature rise is accelerated through the positive-feedback region ($dR/dT > 0$, pink region), resulting in sharp "current-driven" or "first-order-like" transition behavior in the DC *V-I* or *V-t* measurements. Under magnetic fields, the resistance exhibits a more pronounced temperature dependence in this regime, leading to a sharper transition (see Supplemental Fig. 3b & Fig. 5). Above $I_c$, the sample temperature exceeds $T_c$, entering a regime where resistance decreases with temperature ($dR/dT < 0$, orange region). This produces the observed NDR in the *I-V* curves and correlates with the absence of magnetic domains in MFM measurements. We emphasize that analogous Joule heating mechanisms have been implicated in current- or field-driven NDR phenomena in classic transition-metal oxides, such as vanadium, titanium, and niobium oxides[4]. These parallels underscore the universality of electro-thermal feedback in driving abrupt electronic transitions.

Thermal effects typically require longer timescales to reach equilibrium and often induce time delays and hysteretic behaviors in transport measurements. For instance, the time-dependent bistable switching (Y. Zhang *et al.*[27] and Supplemental Fig. 5) with long timescales (seconds to minutes) operates only for currents slightly exceeding $I_c$. This delay arises because the sample requires time to accumulate sufficient energy via Joule heating to enter the positive-feedback regime ($T$>~50 K), where abrupt bistable switching occurs. For the triangular-wave study presented in Fig. 2, the absence of "critical current" like transitions at $f$ >~70 Hz can now be understood as consequence of the sample being heated up above $T_c$ without sufficient time for heat dissipation to the environment. The temperature of the sample at zero current continues to rise with increasing the frequency due to heat accumulation during the sweeping process, resulting in the similarity of the d$V$/d$I$($V$ = 0) versus $f$ curve (Fig. 2e) and the $R$-$T$ curve above 60 K (Fig. 3a). When the required time for thermal equilibrium is much longer than the sweeping rate ($f$ >~1000 Hz), the temperature of the sample no longer varies during the sweeping, exhibiting linear $I$-$V$ and $\eta$ close to 0 again (Fig. 2d). The sample finally reaches a condition with dynamic equilibrium between self-heating and dissipation at $T_s$~120 K (though the environment is still ~60 K).

**Discussion**

In summary, systematic MFM and dynamical transport measurement in $Mn_3Si_2Te_6$ reveal that the Joule heating at high currents plays a critical role in driving the observed current-induced phase transitions and non-reciprocal charge transport. The first-order-like transition triggered by direct currents arises from self-reinforcing heat accumulation via a positive feedback mechanism, enabling the sample temperature to surge by over ~100 K within hundreds of microseconds. Our findings underscore the critical need to account for electro-thermal dynamics when interpreting transport phenomena in quantum materials. While no definitive evidence of COC domains[27] is observed in this study, their potential existence cannot be excluded and necessitates future investigations using advanced techniques with higher magnetic sensitivity and spatial resolution.

**Methods**

**Single-crystal growth**

The $Mn_3Si_2Te_6$ single crystals were synthesized by the chemical vapor transport (CVT) method in two steps. A stoichiometric ratio of manganese powders, silicon powders and tellurium lumps were mixed and sealed in an evacuated quartz ampoule, then slowly heated to 1000 °C and held for one day to get polycrystalline precursors. The products were taken out and ground thoroughly, then resealed in a quartz ampoule with iodine of 5 mg/ml as the transport agent. The ampoule was exposed to a temperature gradient of 800 °C to 750 °C where the CVT preceded for two weeks and was then naturally cooled down to room temperature. $Mn_3Si_2Te_6$ crystals in size of ~5 mm with regular shapes and shiny surfaces were finally obtained. Besides, a slight excess of silicon powders would improve the crystal growth.

**MFM measurement**

The MFM experiments were carried out in a commercial high vacuum cryogenic MFM system, using commercial PPP-MFMR probes fabricated by Nanosensors[TM]. MFM images were taken in a constant height mode with the scanning plane nominally ~ 200 nm above the sample surface. Prior to each MFM scan, a DC bias was applied to the MFM tip to minimize the long-range electrostatic forces between cantilever and sample. During MFM experiments, the resonance frequency shift $\Delta f$ of cantilever was closely tracked by a phase-locked loop and proportional to the second derivative of the stray magnetic field in $Z$ direction. The peaks (dips) in MFM signals represented repulsive (attractive) magnetic force, indicating an anti-parallel (parallel) magnetization direction to tip moments.

**sMIM measurement**

We have conducted magnetic field and current dependent local conductivity measurement by in-situ transport and scanning microwave impedance microscope (sMIM) experiments (Supplemental Fig. 6a). The sMIM experiments were carried out in a commercial cryogenic sMIM system. During sMIM measurement, the cantilever was mechanically excited at its eigenmode. The Pt-Ir tips from Rocky Mountain Nanotechnology LLC were used for non-contact sMIM scanning. The resulting oscillation amplitude of sMIM-Im was then demodulated using the lock-in amplifier to yield d(sMIM-Im)/dz, referred as sMIM signal here. sMIM signals were taken in a constant height mode with the scanning plane nominally ~ 50 nm above the sample surface.

Current dependent sMIM measurement reveals that the local conductivity varies dramatically upon increasing current (Supplemental Fig. 6c). Supplemental Fig. 6d illustrates that local conductivity rapidly increases and then nearly saturates upon increasing the current in region I. Unexpectedly, in region II, local conductivity significantly decreases near $I_c$, suggesting decreased conductivity near the transition. Following this decrease, local conductivity abruptly increases with further current elevation, accompanied by a sharp rise in measured voltage. It is noteworthy that the variation in local conductivity closely resembles the temperature-dependent conductance curve (Supplemental Fig. 6e), supporting a Joule heating-induced phase transition. Furthermore, although the Bloch domain wall in MFM experiments exhibits out-of-plane magnetization, which should be highly conductive and enhance the signal in the sMIM channel, no signature of the domain wall is observed in sMIM experiments. This implies that the Bloch domain wall remains as insulating as the FiM domain.

**Magnetization measurements**

Magnetization measurements were carried out in a commercial 9 T Physical Property

Measurement System (Dynacool-9 T, Quantum design) with 14.2 mg $Mn_3Si_2Te_6$ crystal. Magnetization within ab-plane $M_{ab}$ saturates near 1.6 $\mu_B$/Mn at $\mu_0H \geq 0.05$ T at 2 K, while magnetization along c-axis $M_c$ approaches 1.53 $\mu_B$/Mn at 9 T (Supplemental Fig. 7c).

**Transport measurements**

The transport measurements were performed in a helium-4 vapor flow cryostat with 9 T magnet and temperature range between 1.6 ~ 250 K, which was measured using a Cernox thermometer mounted next to the chip carrier. The surface of $Mn_3Si_2Te_6$ is susceptible to oxidation, which hinders its contact with the electrodes. Ar-ion etching and subsequent metal deposition (Pd/Au ~ 15 nm/85 nm) were carried out to improve electrical contacts. The *R-T* curves were measured using Keithley 6221 and Keithley 2182a with delta mode at 100 μA (upper panel of Supplemental Fig. 3a). Keithley 6221 and Agilent 34410a were used for DC *I-V* characteristic measurement (Supplemental Fig. 3b) and DC *R-T* measurement (lower panel of Supplemental Fig. 3a). Synchronized NF Corporation LI5640 lock-in amplifiers with low frequencies were used to measure magnetoresistance *MR* $(=(\rho_{xx}(H)-\rho_{xx}(0))/\rho_{xx}(0))$ in Supplemental Fig. 3c.

Triangular waveforms with different frequencies $f$ and rectangular waveforms with various pulse current magnitudes $I_a$ were generated by a Keithley 6221 source meter, and a 4-channel 500 MHz oscilloscope DSO-X 3054G with high impedance mode (input resistance 1 MΩ) was used to measure the electrical potential at each contact. The trigger signal output by the source meter guaranteed the synchronized phase of the data acquisition for each measurement cycle.

**Electro-thermal simulation**

The duration of the current pulse is short enough that we can assume the environmental (bath) temperature $T = T_0$ remains constant. We can construct a simplified electro-thermal model that takes into account both heat generation (Joule-heating in sample) and dissipation:

$$nc_P(T_s)dT_s = dQ$$

$$\frac{dQ}{dt} = I^2R(T_s) - K(T_s - T_0)$$

$$K = \alpha T_s^2 + \beta T_s + \gamma$$

, where $c_P(T_s)$ represents specific heat capacity as a function of temperature and adopted from the literature[19,24], $n$ represents the estimated amount of substance of the sample ~1.5×10⁻⁸ mol, *K* describes the interfacial thermal conductance between the sample and its surrounding environment[37] (we use α ~ 2.3×10⁻⁷ W/K³, β ~ 3.2×10⁻⁶ W/K², γ ~ 5×10⁻⁴ W/K), and $R(T_s)$ is the temperature-dependent resistance (see Fig. 3a). In measurements using continuous triangular waveforms, the ambient temperature unavoidably increases. Simulation for higher frequency at higher $T_0$ can qualitatively match the experimental results better, as showed in Supplemental Fig. 4.

We simulated the cross-section view of the temperature distribution near the sample channel with a 40 mA current input and a duration of 500 μs using the finite element method (COMSOL Multiphysics), showed in inset of the main text Fig. 3f. The thermal conductivity, and special heat are functions of the temperature and were adopted from the literature[24,38]. Heat conduction between the sample, silicon (Si) substrates, Au electrodes and copper chip carrier was considered, while heat transfer between the solid interface and the bath environment ($T_0$= 20 K, ~5 Pa He4 vapor) through thermal convection was negligible. The simulation showed the sample could be locally heated by more than ~100 K. Additionally, it revealed a temperature gradient from the top to the bottom of the sample, along with heat transfer to the Si substrates.

**Further discussion for dynamical voltage responses over longer time scales**

Similar time-dependent bistable switching measurements on our sample over longer time scales have been conducted and are shown in Supplemental Fig. 5a and 5b. The results are similar to those previous reports[27]. Here, we clarify that the phenomena can be understood as follows. When a sufficiently large current is applied, it takes a finite amount of time to accumulate enough heat to reach a threshold temperature ~50 K. Beyond this point, the temperature increase can accelerate due to the positive feedback process, resulting in a sharp voltage jump through the positive d$R$/d$T$ region. The larger the current, the shorter the time required for the process to occur. The onset time of voltage jumps follows an exponential way (Supplemental Fig. 5c), expressed as $I = Ae^{-\frac{t}{t_0}} + I_0$, where $A$, $t_0$, and $I_0$ are fitting parameters. This expression also supports the Joule-heating mechanism[10]. This process can be viewed as dynamical responses at smaller currents and longer time scales compared to that measured in Fig. 3.

**Acknowledgements:**

This work was funded by National Natural Science Foundation of China (Grants No. 12174439, 62488201, 12374199), the Innovation Program for Quantum Science and Technology (Grant Nos. 2021ZD0302400 and 2021ZD0302300). Crystal growth by Z. Li was supported by the Youth Innovation Promotion Association of CAS (No. 2021008).

**Figures**

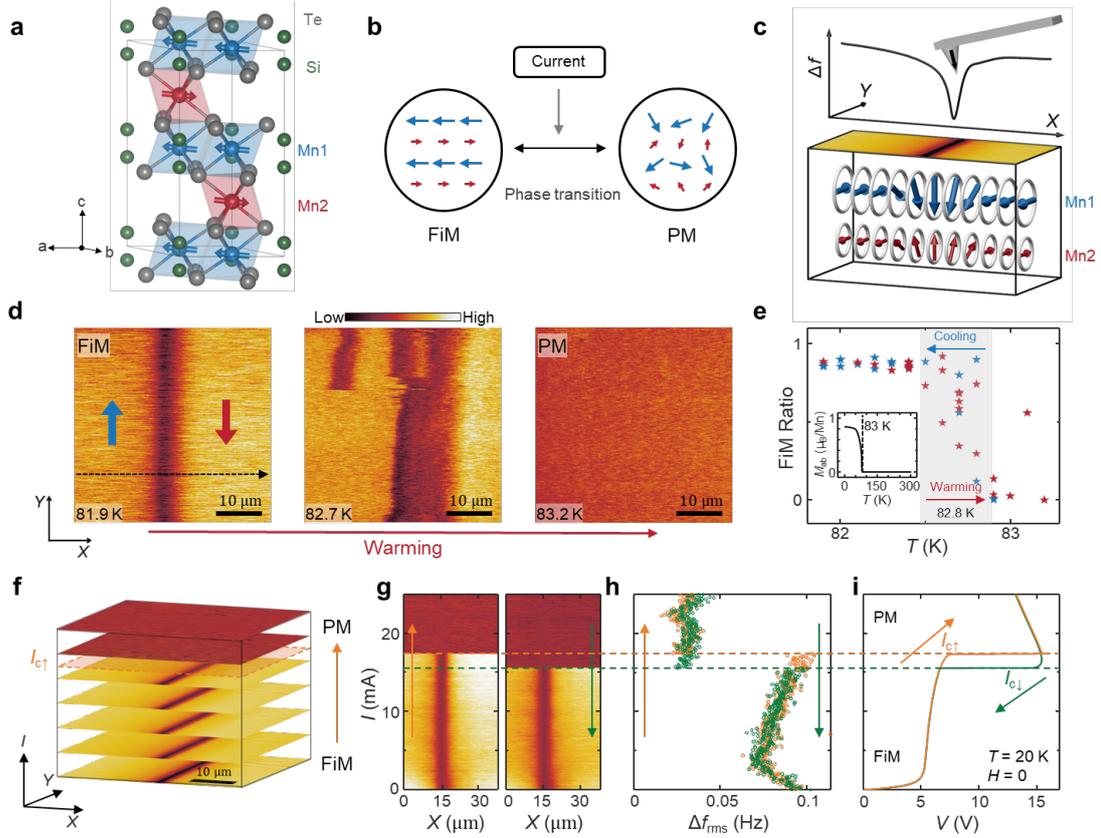

**Fig.1 | Visualization of magnetic ground state of Mn$_3$Si$_2$Te$_6$ by magnetic force microscope (MFM). a**, Crystal and magnetic structure of Mn$_3$Si$_2$Te$_6$ below the ferrimagnetic transition temperature $T_c$. **b,** Schematic of current induced magnetic phase transition in Mn$_3$Si$_2$Te$_6$. **c**, Average MFM line profile (top), MFM mapping (43×43 µm$^2$ at $T$ = 25 K) and schematic of spin structure across the FiM domain wall (bottom). The in-plane spins in the FiM region continuously rotate to the out-of-plane direction at a Bloch-type domain wall, generating an attractive magnetic force and a negative frequency shift of the MFM cantilever resonance frequency. The color scale is 1.2 Hz. **d**, Temperature-dependent MFM mapping. As sample warmed from 81.9 K (below $T_c$) to 83.2 K (above $T_c$), the FiM domain wall expands and evolves to a PM phase. **e**, The FiM region ratio extracted from MFM images measured at different temperatures across $T_c$. Inset: temperature-dependence of the $ab$-plane magnetization $M_{ab}$ at $\mu_0 H$ = 10 mT, showing FiM-PM transition at $T_c$ = 83 K. **f**, Stack of MFM images from $I$ = 0 mA to 17.5 mA at $T$ = 20 K. As current increases, the FiM phase persists at $I < I_{c\uparrow}$ and transitions to the PM phase abruptly at $I_{c\uparrow}$. The color scales vary slightly for clarity. **g,** Repeated line scans (along the black dashed arrow in **d**) performed while the current is varied (y-axis). **h,** Current dependence of the root-mean-square value $\Delta f_{rms}$ of MFM linecuts during the up-swept (orange) and down-swept (green) process. **i,** Hysteretic $I$-$V$ curves at $T$ = 20 K, showing sharp transitions at critical currents $I_{c\uparrow}$ = 17.4 mA (sweeping up) and $I_{c\downarrow}$ = 15.6 mA (sweeping down). The voltage $V$ and MFM linecuts are recorded simultaneously in **g** and **i**.

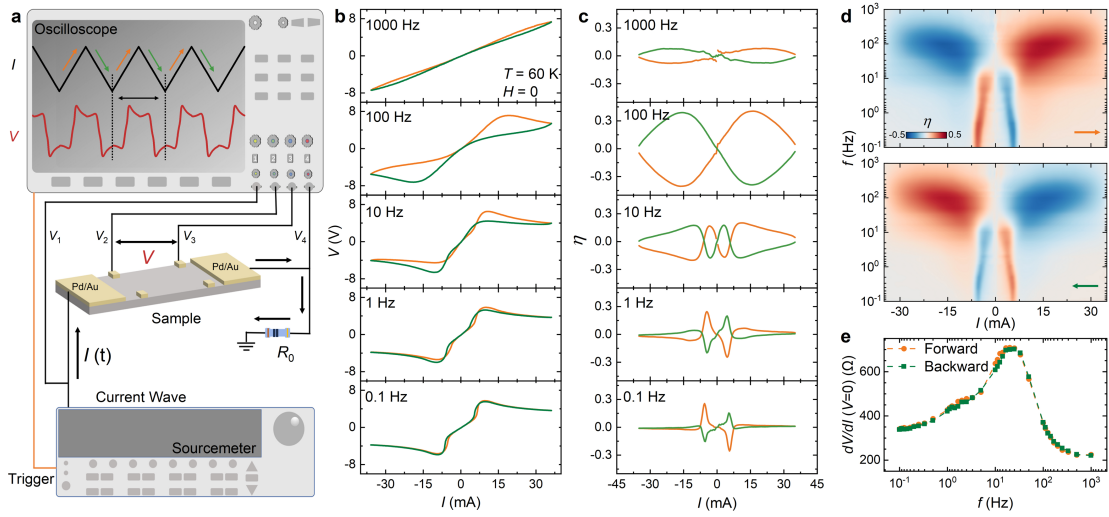

**Fig.2 | Dynamical voltage responses with supplying triangular waveforms. a**, Experimental setup of the dynamic *I-V* measurement. Triangle waves with frequency *f* are generated by a current source meter and a 4-channel oscilloscope is used to measure the electrical potential $V_i$ (i=1, 2, 3, and 4) at each lead of the sample. The four-terminal voltage $V (= V_2 - V_3)$ is obtained by subtracting the signal of Channel 3 from Channel 2. A resistor of $R_0 = 10\ \Omega$ is used to measure the real-time current $I = V_4/R_0$. A representative oscillogram of *I* and *V* is shown on the screen of the oscilloscope. **b**, *I-V* characteristics of five different frequencies (1000, 100, 10, 1, and 0.1 Hz) at 60 K, with the forward (orange lines) and backward (green lines) scan directions of the current denoted by orange and green arrows in **a**, respectively. **c**, Bi-directional nonreciprocity $\eta = \frac{R(I)-R(-I)}{R(I)+R(-I)}$ calculated for each *I-V* curve at frequencies and sweeping directions (color coded) corresponding to **b**. **d**, Color mapping of the bi-directional $\eta$ extracted from the *I-V* characteristics at various frequencies. **e**, The evolution of zero-bias differential resistance (d*V*/d*I*) extracted from the *I-V* curves with changing the frequency.

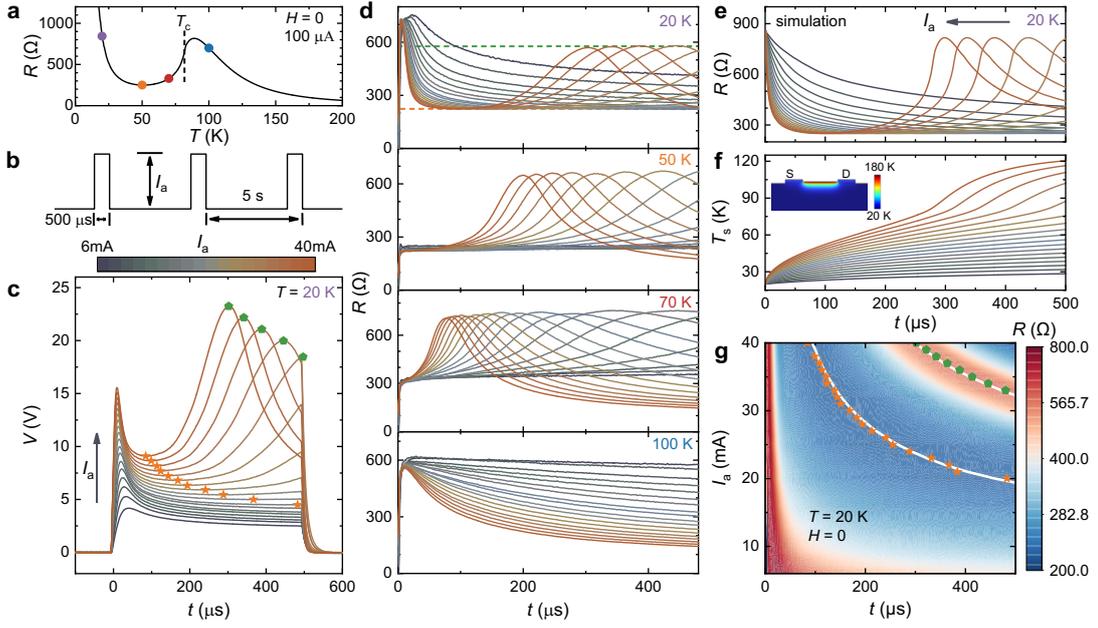

**Fig.3 | Dynamical voltage response with supplying rectangular waveforms. a**, Resistance as a function of the temperature measured with a small DC $I$ = 100 μA. **b**, Schematic diagram of the repetitive current pulses with duration ~ 500 μs and periodicity ~ 5 s. **c**, Voltage response at various pulse amplitudes ($I_a$) as a function of time at 20 K. The curves are color-coded according to the legend bar and take intervals of 2 mA. The arrow indicates the direction of increasing $I_a$. **d**, Resistance $R$ (= $V/I$) as a function of time with increasing $I_a$ (grey to orange curves) at 20 K, 50 K, 70 K, 100 K, corresponding to the purple, orange, red, and blue dots in **a**, respectively. **e-f**, Time-dependent resistance and sample temperature ($T_s$) with varying pulse amplitudes at 20 K, simulated from a simplified Joule-heating model described in the main text. The inset graph is the cross-sectional diagram showing the temperature gradient across the sample (placed on a thick Si substrate) after applying a 40 mA current for a duration of ~ 500 μs, simulated by COMSOL. S and D stand for source and drain, respectively. **g**, Color mapping of the resistance as a function of $t$ and $I_a$ (same data as the first panel of **d**), with the corresponding dips and peaks in $R(t)$ marked by the orange stars and green pentagons, respectively. The dip and peak positions extracted from the simulation are illustrated as white lines.

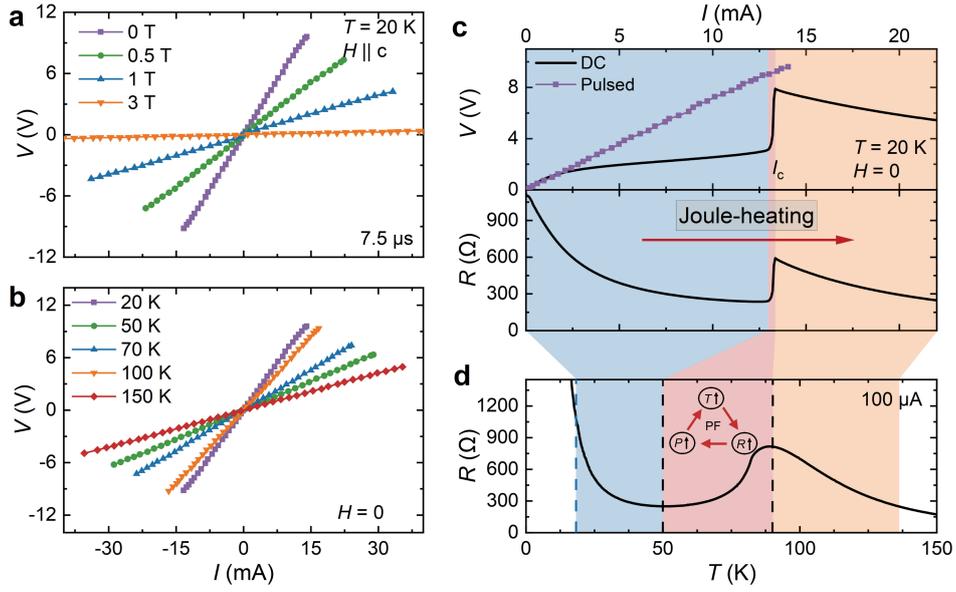

**Fig.4 | Intrinsic *I-V* characteristics and the Joule-heating mechanism. a-b,** *I-V* characteristic extracted from pulsed-current measurement (delay time 7.5 μs) with various magnetic fields at *T* = 20 K (**a**) and with various temperatures at *H* = 0 (**b**). **c**, Upper panel: comparison of *I-V* curves from the DC measurement (in black) and pulsed-current measurement (in purple) at 20 K and *H* = 0. Lower panel: resistance (calculated by *V*/*I* from the DC measurement) as a function of the current. **d**, Replotted *R-T* curve (same data as Fig. 3a), strongly resembling the *R-I* curve. The color blocks serve as guides for corresponding segments in each curve. Inset: Diagram of a positive feedback (PF) process for the transition region.